\documentclass{iau} 
\usepackage{graphicx}
\usepackage{natbib}
\usepackage{hyperref}

\def\fw{\textsc{fastwind} }
\def\phoebe{\textsc{phoebe} }
\def\spamms{\textsc{spamms} }

\title[SPAMMS: applications and use cases] 
{SPAMMS: applications and use cases for the 3D spectroscopic analysis technique to study deformed massive stars}

\author[M. Abdul-Masih]   
{Michael Abdul-Masih$^{1}$}

\affiliation{$^1$European Southern Observatory, Alonso de Cordova 3107, Vitacura, Casilla 19001, Santiago de Chile, Chile
}

\pubyear{2022}
\volume{361}  
\jname{Massive Stars Near and Far}
\editors{N.~St-Louis, J.~S.~Vink \& J.~Mackey, eds.}
\begin{document}

\maketitle

\begin{abstract}
Whether it be due to rapid rotation or binary interactions, deviations from spherical symmetry are common in massive stars.  These deviations from spherical symmetry are known to cause non-uniform distributions of various parameters across the surface including temperature, which can drive internal mixing processes within the envelopes of these massive stars.  Despite how common these 3D distortions are, they are often neglected in spectroscopic analyses.  We present a new spectral analysis code called \spamms (Spectroscopic PAtch Model for Massive Stars) specifically designed to analyze non-spherical systems.  We discuss how the code works and discuss its assumptions.  Furthermore, we demonstrate how \spamms can be applied to a variety of different types of systems and we show how it can model 3D effects in a way that current analysis techniques are not able to.

\end{abstract}

\firstsection 
\section{Introduction}

Deviations from spherical symmetry are common throughout the lifetime of massive stars \citep{Langer2012}. Nearly a quarter of all massive stars will merge with a companion during their lifetimes, and this coalescence is preceded by a highly non-spherical contact phase \citep{Pols1994, Sana2012}. The internal processes occurring during the contact phase can have large effects on the eventual end-products of these systems \citep[e.g., ][]{Shao2014, deMink2016, Mandel2016, Marchant2016, Schneider2019}. Even in massive single stars, deviations from spherical symmetry are common due to rotational distortion \citep{Ramirez-Agudelo2015, Bodensteiner2020b}. The three-dimensional geometry of these non-spherical systems is, however, typically not taken into account in conventional tools available for the spectroscopic analysis of massive stars.

Due to the temperature regime where massive stars are found, NLTE effects become very important to accurately model the spectral lines.  Thus, radiative transfer codes suitable for massive stars are typically computed in 1D because the NLTE calculations are computationally expensive \citep[e.g., ][]{Hillier1998, Grafener2002, Puls2005}.  This imposes the assumption of spherical symmetry on the system, and ignores any effects arising from the surface variations resulting from the 3D geometry.  Here, we present a new spectral analysis code called /textsc{spamms}, which is specifically designed to analyze these types of distorted systems, and we demonstrate its ability to handle various geometries. We then apply it to a sample of massive overcontact binaries and compare its performance to traditional spectroscopic binary analysis techniques. Additionally, we demonstrate how the code can be used to investigate the degree to which the 3D geometry affects the observed temperature in rapidly rotating systems.  Finally, we show how the code can model various 3D effects that occur in spherical binary systems as well.

\section{SPAMMS} \label{sec:spamms}
\begin{figure}
\begin{center}
 \includegraphics[width=\textwidth]{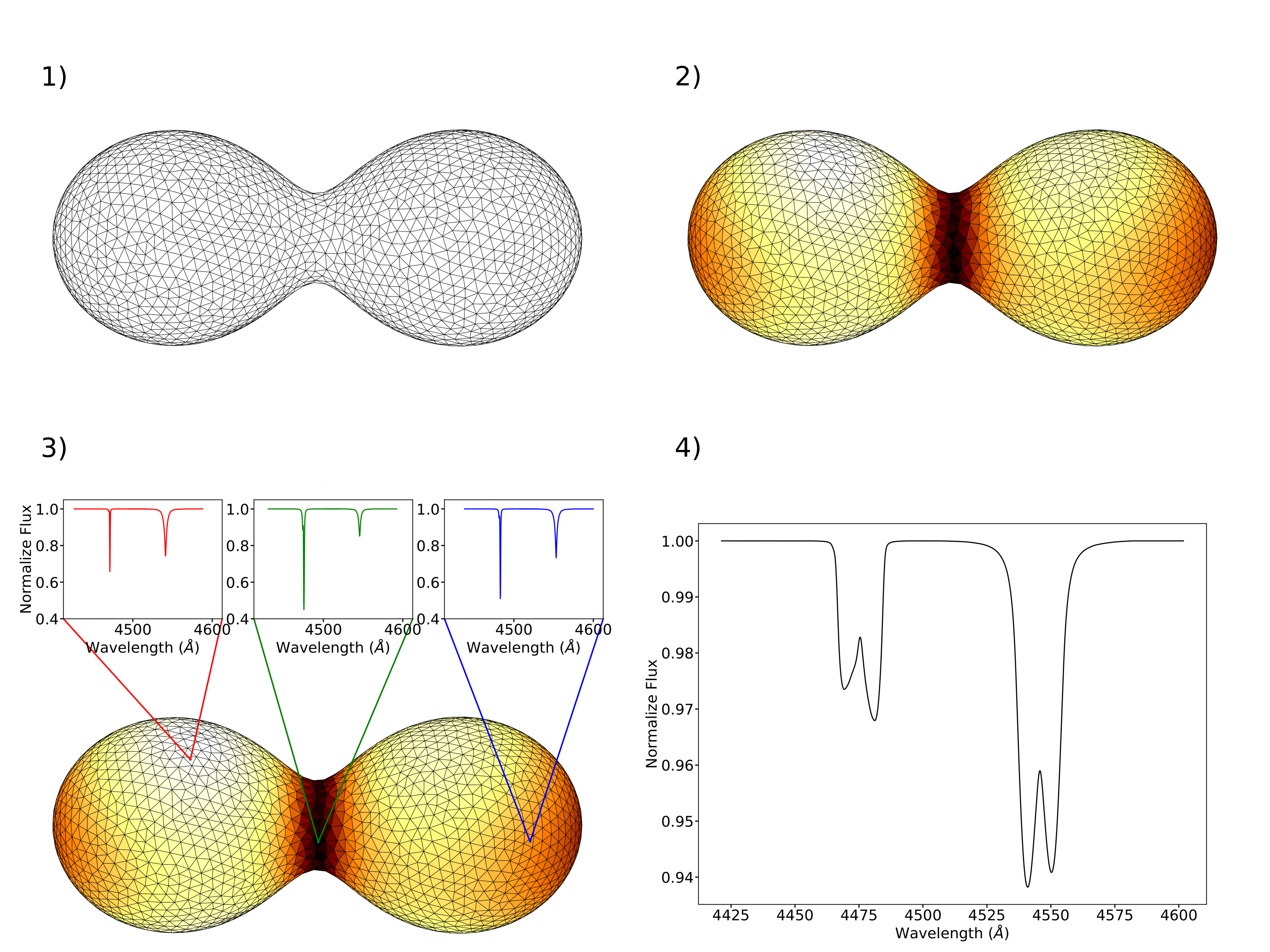} 
 \caption{Graphical representation of the four step \spamms model computation as described in Sec. \ref{sec:spamms}.  1) represents the generation of the mesh. 2) represents the population of the mesh with local parameters.  Here the surface temperature distribution is shown as the face colors of each patch triangle. 3) represents the assigning of \fw models to each patch. 4) shows the final integrated line profile for the given system at the given phase. This figure originally appeared in \citet{Abdul-Masih2020b}}
   \label{fig:spamms}
\end{center}
\end{figure}

In order to properly model distorted massive stars, we require a code that can simultaneously model the 3D geometry (and resulting surface parameter distributions) and model the NLTE effects in the atmosphere.  For this purpose, we present the Spectroscopic PAtch Model for Massive Stars \citep[\textsc{spamms}; ][]{Abdul-Masih2020b}.  This code combines the 3D modelling capabilities of \phoebe \citep{Prsa2016} with the NLTE radiative transfer code \fw \citep{Puls2005} to create a spectroscopic patch model across the surface of a three dimensional system.  

\spamms works in 4 steps.  First, given the physical stellar or binary parameters of the system (radius, mass, inclination, rotation rate, period, phase, etc.), a 3D triangulated mesh is generated accounting for Roche or rotational distortions.  Next, the mesh is populated with local parameters such as temperature, surface gravity, radius and radial velocity.  Third, based on the local parameters, \fw synthetic line profiles are assigned to each patch across the surface.  Finally, the visible surface is integrated over, accounting for surface area projection effects, emergent angles and the intensity contributions from each patch.  This results in a final synthetic line profile that accounts for the geometry and the surface effects of the system.  A graphical representation of these steps can be found in Fig. \ref{fig:spamms}.

\section{Applications}

\subsection{Overcontact binaries}
The overcontact phase is expected to occur in about a quarter of all massive stars \citep{Sana2012} and is characterized by a highly distorted geometry \citep[see e.g., ][]{Fabry2022}.  This evolutionary phase is of particular importance in the context of mergers and gravitational wave progenitors, so an accurate determination of surface parameters can be vitally important in understanding the future evolution of these systems.  Traditional fitting techniques involve first disentangling the observed spectra and then fitting each component with 1D models \citep[e.g., ][]{Abdul-Masih2019}.  This imposes the spherical assumption several times throughout the process and can bias the derived stellar parameters.  

To investigate how big of a difference accounting for the 3D geometry has on the derived solution, we modeled a sample of three massive overcontact systems using both \spamms and traditional fitting techniques. Since the three overcontact systems had been characterized photometrically in the past \citep{Hilditch2005, Almeida2015, Martins2017}, we use the binary solution from these works as the input for the \spamms model to constrain the geometry. As shown in \citet{Abdul-Masih2021}, we are able to reproduce the observed spectra to a higher accuracy with significantly fewer free parameters using \spamms than with the traditional fitting techniques (8 and 22 free parameters respectively).

\begin{figure}
\begin{center}
 \includegraphics[width=\textwidth]{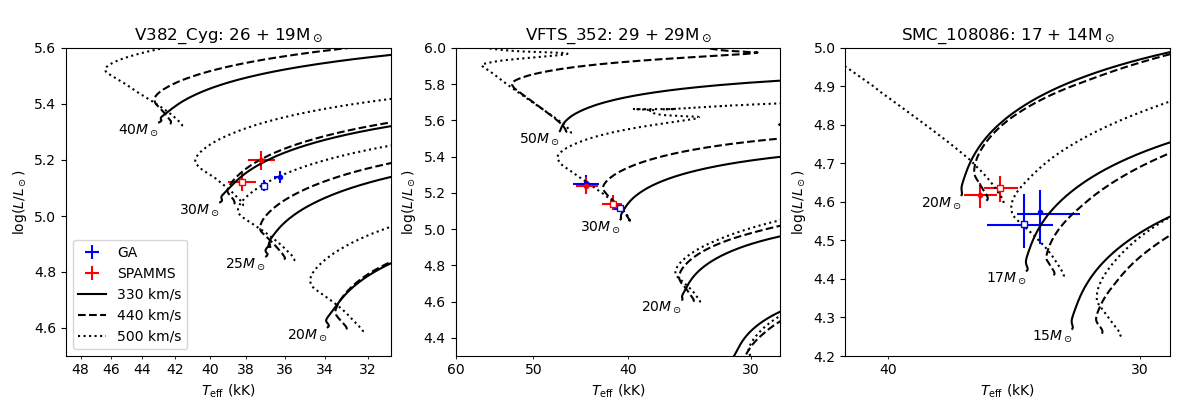} 
 \caption{HRDs for the three overcontact systems that we analyzed.  The positions of the primary and secondary are indicated with closed and open points respectively.  The \spamms fit is indicated in red while the fit from the 1D techniques are indicated in blue.  For reference, rotating evolutionary tracks for various masses from \citep{Brott2011b} are plotted.  This figure is adapted from \citet{Abdul-Masih2021}}
   \label{fig:overcontacts}
\end{center}
\end{figure}

In addition to better morphological agreements between the observed spectra and the models, we also found a difference in the derived parameters using both techniques.  The most significant and important of these differences involves the measured temperature and luminosities.  We find that for the systems with inclinations close to 90, \spamms measures a higher temperature and luminosity than the 1D techniques.  On the other hand, for the system with an inclination close to 60, the measured temperatures and luminosities agree quite well.  This can be seen in Fig. \ref{fig:overcontacts}. This is significant, because the difference in the measured temperatures and luminosities has a large impact on our understanding of the internal mixing processes for these systems, and thus their future evolution.

\subsection{Rapid rotators}
Motivated by our initial benchmarking of \spamms \citep[see Sec. 3 of ][]{Abdul-Masih2020b} as well as the differences in measured temperature found for the overcontact systems when using \spamms and 1D techniques, we investigated how the geometry of rapidly rotating systems affects the stellar parameters.  As with overcontact systems, the geometry of rapidly rotating systems can deviate quite strongly from spherical symmetry, and the temperature range across the surface is expected to be more extreme for systems approaching critical rotation than in the case of overcontact systems.

Using \textsc{spamms}, we generate synthetic spectra for a system rotating at 90\% of critical velocity at different inclinations. We then perform a Monte Carlo simulation where we simulate a signal to noise of 300 and fit the synthetic \spamms spectra with 1D models to characterize how the measured temperature varies with inclination.  We find that at most inclinations, the measured 1D temperature is more than $3\sigma$ away from the input temperature.  At lower inclinations, a hotter temperature is measured, while at higher inclinations a coolet temperature is measured, with equivalent temperatures being found at an inclination of $\sim 60^{\circ}$.  Our findings indicate that not only does the temperature change by up to 5000 K based on the inclination, but also that the measured helium abundance also changes by a significant amount.  

\subsection{Binary effects}
In addition to the two classes of objects discussed above, \spamms has also been applied to semi-detached and fully detached binary systems.  In both of these cases, we find that \spamms is able to reproduce interesting 3D effects.

One such effect is known as the Struve-Sahade effect \citep{Struve1950, Sahade1959, Linder2007}.  This effect is characterized by a strengthening of spectral lines when they are blue-shifted and a weakening when they are red-shifted.  This effect is typically associated with semi-detached binaries and work by \citet{Palate2012} demonstrated that this effect could be reproduced using a similar approach to \spamms.  In \citet{Abdul-Masih2020b}, we demonstrated that \spamms can also reproduce this effect, however in this case, we saw it when modeling an overcontact system.  It is still unclear what causes this effect, but the fact that it can be produced by codes such as \spamms indicate that at least some of the effect is geometrical in nature.

\begin{figure}
\begin{center}
 \includegraphics[width=\textwidth]{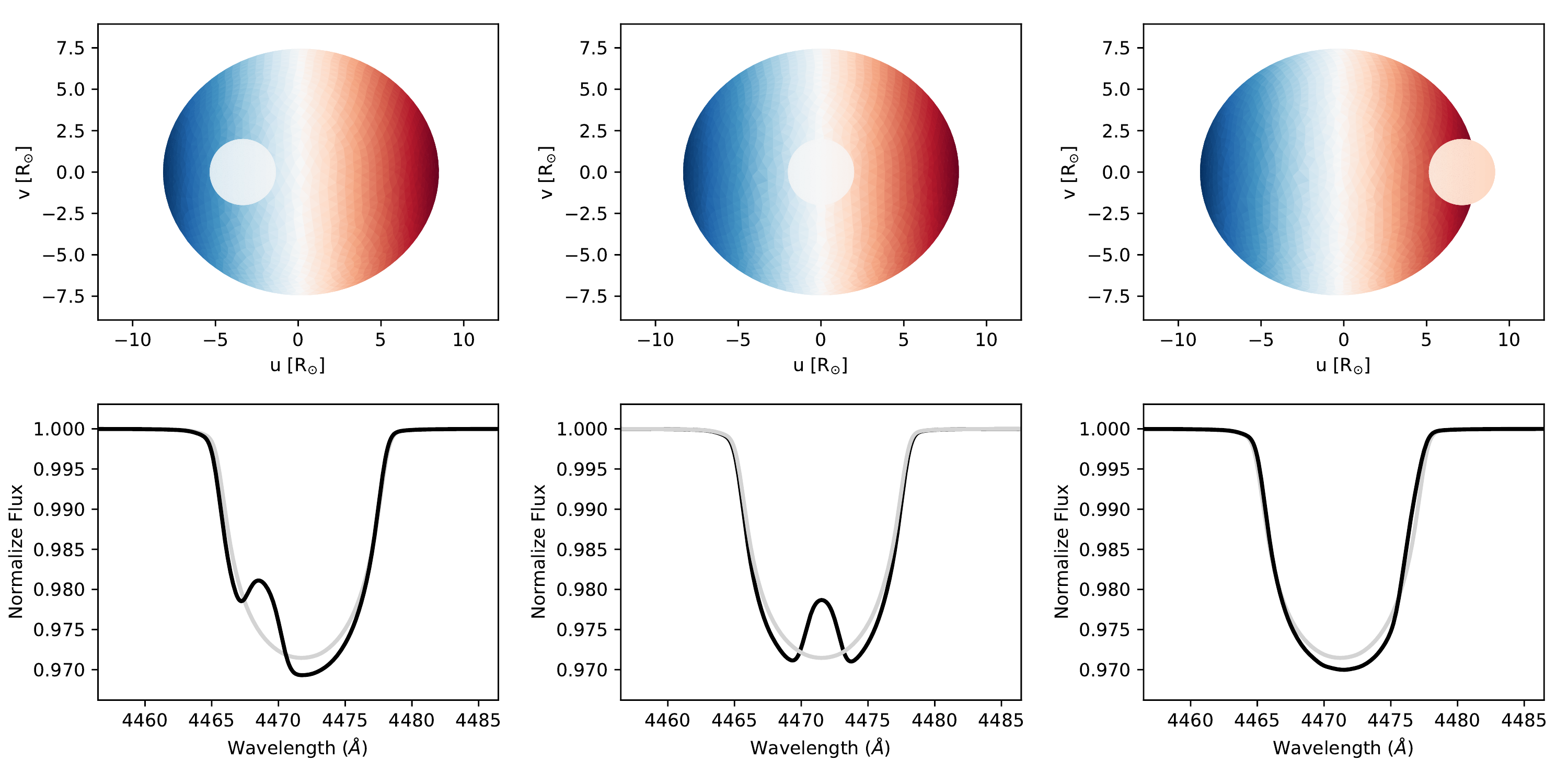} 
 \caption{Example of the Rossiter Mclaughlin effect.  The top panels show a rotating star at various phases during the eclipse with the face color representing the radial velocity profile across the surface.  The bottom panels show how the line profile changes during the eclipse with the out of eclipse line profile plotted in grey for reference.  This figure originally appeared in \citet{Abdul-Masih2020b}}
   \label{fig:rossiter}
\end{center}
\end{figure}

Another effect that \spamms can reproduce is known as the Rossiter Mclaughlin effect \citep{Rossiter1924, Mclaughlin1924}.  This effect occurs when a rotating star is eclipsed by a companion. Since the rotating star has regions of the surface that are red-shifted and others that are blue-shifted, when the star is eclipsed by its companion, the spectral lines of the rotating system get skewed in different ways depending on which region of the star is being eclipsed.  This is demonstrated in Fig. \ref{fig:rossiter}  This effect is often observed when a planet eclipses its host star and is very important for determining if there is a spin-orbit misalignment.

\section{Conclusions and future prospects}
Here, we have presented \textsc{spamms}, a spectroscopic analysis code specifically designed for distorted stars.  We have explained how the code works, and how it is able to reproduce the 3D effects associated with non-spherical geometries.  We then described various ways in which \spamms has been applied, and we have shown some of the results to have come from this code so far.  

While we have gone through some examples, this is by no means the limit of what \spamms can model.  In addition to the cases shown, \spamms can spectroscopically model anything that \phoebe can create a mesh for.  This means that \spamms benefits from all of the advanced physical prescriptions and phenomena that are already included in \phoebe including things such as reflection effects, spots, Doppler boosting, and many more.  Additionally, as \phoebe evolves and includes more features, so will \textsc{spamms}.  Some notable features that are currently in development that will be particularly interesting for \spamms include pulsations, alternative gravity darkening prescriptions and the ability to model higher order multiple system.  

\spamms represents an exciting step forward in our understanding and ability to model massive star systems.  With each new \spamms release, we will be able to model new and different types of systems that up until now we have not had the ability to model in 3D.

\bibliographystyle{apj}
\bibliography{mybib}

\begin{discussion}

\discuss{F. Martins}{
  You showed for rapidly rotating stars that these effects are important.  I was wondering if you can you say how close binaries need to be for these effects to be important
}

\discuss{M. Abdul-Masih}{
  These effects arise from the temperature distribution across the surface, so in general, the more deformed your system is, the larger these effects are.  I would say that the deformations due to rotation and the Roche geometry are more important than how physically close the two stars are, so I can't really cite a specific value for this, but we do plan to investigate what level of distortion is needed for these effects to become important in binary systems.
}

\discuss{A. Herrero}{
  I have two short questions, the first is concerning the surface gravity.  Do you see the same effects on the surface gravity that you do on the temperatures and if so what is the range? And what is the effect when you change the temperature, do you see the same effects at lower and higher temperatures?
}

\discuss{M. Abdul-Masih}{
  We do see the same effect for surface gravities, you would see lower surface gravities when looking at an inclination of 90.  The change in surface gravity with inclination depends quite heavily on how rapidly rotating your star is, but at close to critical, you can have differences in $\log{g}$ of $\sim 0.5$ or more.  
  
  When we change the temperature we still see the same inclination based effect.  In this study, we focused on helium lines as temperature diagnostics, but as you shift to higher or lower temperatures you can use the ionization balance of other species to do the same thing.  
}
\end{discussion}

\end{document}